# Halogen adsorption and reaction with Bi$_2$(Se,Te)$_3$ and Bi/Bi$_2$(Se,Te)$_3$

Haoshan Zhu, Weimin Zhou and Jory A. Yarmoff*

*Department of Physics and Astronomy, University of California, Riverside, Riverside, CA 92521*

**Abstract**

Bi$_2$Se$_3$ and Bi$_2$Te$_3$, and these same surfaces covered with Bi films, are exposed to Br$_2$ and Cl$_2$ in ultra-high vacuum. Low energy electron diffraction (LEED) and low energy ion scattering (LEIS) are used to investigate the surface composition before and after halogen exposure. It is found that Br$_2$ weakly chemisorbs to the Se- or Te-terminated clean surfaces and light annealing removes the adsorbates restoring the intact surfaces. In contrast, halogens dissociatively adsorb onto surfaces covered with an additional bilayer of Bi, having a p-doping effect. Annealing these halogen-covered surfaces at 130ºC causes Bi atoms to be chemically etched away and the surface reverts to a Se- or Te-termination. This work shows how halogen adsorption and reaction can be used to modify the surface termination of such materials.

*Corresponding author, E-mail: yarmoff@ucr.edu

# I. Introduction

Over the past decades, the fundamental science underlying halogen adsorption and reaction with semiconductor materials, such as silicon [1,2] and III-V compound semiconductors [3-5], has been extensively studied because of the utility of these reactions in modifying surfaces and in forming device structures, typically through reactive ion etching (RIE) in a halogen containing plasma [6,7]. Halogens are also promising candidate molecules for atomic layer etching, which is one of the basic techniques being developed to fabricate nanometer-scale structures [8,9]. If the gaseous halogen reactant is in the stable molecular form, then the halogen-halogen bond must be broken at a reactive surface site before adsorption can occur, which means that halogens do not necessarily adsorb and react with all materials [10]. The adsorption of halogens on surfaces is thus complex and related to the surface termination, atomic structure and number of surface defects.

$Bi_2Se_3$ and $Bi_2Te_3$ are topological insulators (TIs) that have attracted intensive interest in recent years. TIs are promising candidates for superconductor, spintronics and quantum computing applications [11-13]. $Bi_2Se_3$ and $Bi_2Te_3$ single crystals are compound two-dimensional (2D) materials composed of weakly interacting quintuple layers (QL) that are arranged as Se-Bi-Se-Bi-Se or Te-Bi-Te-Bi-Te, respectively. The stable surfaces are obtained in a clean environment by cleavage between QLs, and are thus terminated with either Se or Te.

The reactions of $Bi_2(Se,Te)_3$ with several gases have been studied extensively because the band structure has been observed to decay over time, the so-called aging effect, which is possibly due to reaction with atmospheric contaminants leading to the formation of surface Se or Te vacancy defects [14,15]. Thus, many investigations have focused on the reactions of $Bi_2(Se,Te)_3$ with $O_2$ and $H_2O$ [16-19]. $NO_2$ exposure has also been used to dope the surface and tune the Fermi level of $Bi_2Se_3$ to coincide with the Dirac point, which is the optimal position for device applications. Using a gas to dope the surface could tune the Fermi level position while not destroying the topological electronic states, as they are protected from nonmagnetic impurities [18,20]. The reaction of $Bi_2Se_3$ with $H_2$, which etches surface Se and leaves a Bi bilayer (BL) film on surface, has also been investigated [21]. The reactions of halogens with $Bi_2(Se,Te)_3$ surfaces have not yet been reported, however.

Such reactions may be comparable to halogen reaction with III-V semiconductors in which different surface terminations result in different adsorption behavior [3,22-25]. For example, previous work from our group showed that halogens adsorb on III-V semiconductors that are terminated with the electron deficient group III atoms and that post-annealing causes the thermal desorption of group III halides, which leads to the surfaces becoming group V-terminated. Halogen reaction with group V-terminated materials leads to a disordered surface, as the halogens adsorb by breaking bonds between the upper two layers and attaching to second layer group III atoms. The preferential reactivity of halogens with particular atoms in a compound material leads to the possibility of selective chemistry in which certain elements can be etched from the surface leading to methods for intentionally altering the termination and surface structure.

The termination of clean TI surfaces has been studied extensively because of its contribution in determining the surface topological states [26-28]. $Bi_2(Se,Te)_3$ surfaces cleaved in ultra-high vacuum (UHV) and those prepared by ion bombardment and annealing (IBA) are naturally terminated with Se or Te [26,29,30], but theory and experiment have also shown that a surface terminated with an additional Bi BL is particularly stable and energetically favorable compared to the pure QL layer-terminated surfaces [31-34]. Other work has shown that cleaving surfaces in air, or subsequent exposure to air, can lead to a Bi-termination [16,27]. It is thus interesting and important to explore the adsorption and reaction of halogens with TI surfaces that have different terminations.

Ultrathin Bi films on TI substrates have also attracted interest for their nontrivial properties such as a large magnetoresistance and topological edge states [35,36], and an isolated Bi(0001) bilayer (BL) is also a 2D topological insulator [37,38]. Hetero-structure engineering with Bi BLs and other materials provide the ability to control the electronic properties and fabricate unique device structures [39-43]. Among them, bottom-up methods such as molecular beam epitaxy (MBE) are the most commonly used fabrication techniques, although a top-down approach has also been reported [21]. Bi can react with halogen gases directly as does Si, which provides the possibility of etching Bi in a top-down process and forming a nanoscale device structure. This provides a further motivation for exploring halogen reaction with TIs and Bi-covered TIs.

The present paper presents a low energy electron diffraction (LEED) and low energy ion scattering (LEIS) study of halogen adsorption and etching of pure $Bi_2(Se,Te)_3$ and Bi-covered $Bi_2(Se,Te)_3$ surfaces. It is found that some molecular halogens adsorb on clean Se- or Te-terminated $Bi_2(Se,Te)_3$ surfaces, but the materials are not etched by the halogens. In contrast, surfaces covered by Bi thin films are more reactive to halogens and the Bi films are etched away by heating at a relatively low temperature.

## II. Experimental procedure

Single crystal $Bi_2Se_3$ and $Bi_2Te_3$ boules are grown using a multi-step heating method [44] following the procedures described previously [26,45]. The material cleaves easily along the (0001) plane producing approximately 5 mm diameter samples. The samples are attached to a transferable tantalum (Ta) sample holder for measurements using spot-welded Ta strips.

The surface analysis measurements are performed in an ultra-high vacuum (UHV) main chamber with a base pressure of $2\times10^{-10}$ Torr or below. This main chamber contains a sputter gun for sample cleaning, solid-state electrochemical cells for halogen exposure [46], optics for LEED, and the equipment needed for LEIS that is described below. The sample holders are transferred from a load-lock chamber onto a manipulator that allows for positioning in the x-y-z directions and rotations about both the polar and azimuthal axes. The surfaces are heated radiatively using an e-beam filament mounted behind the sample without a bias voltage, as the annealing temperatures needed to prepare clean and ordered $Bi_2(Se,Te)_3$ surfaces are rather low. The sample is held at room temperature during halogen exposures and spectra collection.

Ion bombardment and annealing (IBA) is used to prepare clean and well-ordered surfaces [29]. For preparing Se-terminated $Bi_2Se_3(0001)$, the IBA process involves 30 min of 500 eV $Ar^+$ ion bombardment at a current density of $2.5\times10^{12}$ $cm^{-2}$ $sec^{-1}$ with the sample at room temperature, followed by annealing in UHV at 490°C for 30 min [26]. Te-terminated $Bi_2Te_3(0001)$ surfaces are produced in the same way except that the annealing temperature is 340°C.

A MBE system is attached to the main chamber in a manner that enables samples to be transferred between the chambers while remaining under UHV. This MBE system is used for growing Bi films onto the $Bi_2(Se,Te)_3$ surfaces, which are held at room temperature during growth. Bi is evaporated at a rate of 1.45 Å $min^{-1}$ from a Knudsen cell heated to 530°C, with the rate being calibrated by a quartz crystal microbalance (QCM). The coverages of Bi are reported here in units of bilayers.

Low energy ion scattering (LEIS) is performed in the main UHV chamber. The ions are produced from a thermionic emission $Na^+$ ion gun (Kimball Physics) that can be operated in a constant current mode or can produce a pulsed beam of ions. The scattered particles are collected by either of two detectors, as described below. The gun is mounted on a rotatable turntable that allows for the scattering geometry to be adjusted. For the present data, the incident ion kinetic energy is fixed at 3.0 keV.

The first detector is used for time-of-flight (TOF) spectroscopy [47]. The beam is pulsed at 100 kHz and incident normal to the sample surface, while the scattered projectiles are collected at scattering angles of 125° for $Bi_2Se_3$ and 130° for $Bi_2Te_3$. The detector consists of a triple microchannel plate (MCP) array located at the end of a 0.57 m long flight tube. The entrance to the MCP detector is held at ground potential so that the sensitivity to ions and neutrals are equal. A bias voltage of 400 V applied to deflection plates in the flight tube separates the scattered ions from the neutrals. Ground potential and 400 V are applied interchangeably to enable the simultaneous collection of separate spectra for the total scattered yield and the scattered neutrals. The MCP detection sensitivity decreases rapidly as the kinetic energy of the scattered projectiles falls below about 2 keV [48], which is part of the reason that a second detector is employed.

The second detector is a Comstock electrostatic analyzer (ESA), operated in constant pass energy mode, that collects only the scattered ions [49]. It is used here for detecting the presence of lighter elements, in particular Cl, as the TOF setup does not allow for a scattering angle less than 100° at which Na scatters from Cl at a kinetic energy that is below the detection limit of the MCP. Instead, the ESA accelerates the charged projectiles before they impact the internal dual MCP array so that there is no low energy cutoff. For the ESA data collected here, the gun is set to provide a scattering angle of 40°, which enables the detection of adsorbed Cl. The beam is incident at a 70° angle from the surface normal when making these measurements.

Bromine and chlorine molecules are produced from solid-state electrochemical cells based on a silver bromide or chloride pellet, respectively, which are mounted on silver foils [22,46]. The cells are heated to approximately 120°C to enable ionic conduction through the pellet and operated at a current of about 10 μA. The exposures are reported here in terms of the cell current integrated over time and given in units of μA hrs. Spectra are collected immediately after each halogen exposure. Samples are also annealed at 130°C after halogen adsorption to explore etching of the material. The exposure or annealing time is varied for different halogens and sample materials to ensure saturation or complete reaction, which is identified when there are no further changes in the spectra with additional exposure or annealing.

**III. Results**

$Bi_2Se_3$, $Bi_2Te_3$ surfaces and Bi BL-capped $Bi_2(Se,Te)_3$ surfaces are exposed to bromine and chlorine molecules to study the reactions of halogens with TI surfaces. Section III.A describes how LEIS is used to obtain information from these materials using the clean surfaces as examples. The results for halogen exposure and post-annealing for clean Se- and Te-terminated surfaces are presented in section III.B. Section III.C provides results for halogen adsorption onto the TI surfaces when they are covered with a Bi film. Section III.D explores the chemical reactions that occur when the halogen-covered Bi-terminated surfaces are annealed.

**A. Low Energy Ion Scattering from TI surfaces**

LEIS spectra are used to investigate the surface composition and structure. As an example, the upper panels of Fig. 1 show total yield (ions plus neutrals) TOF spectra collected from clean Te-terminated $Bi_2Te_3$. In LEIS, the target can be considered to consist of unbound atoms located at the lattice sites, as the bonding energy is much smaller than the kinetic energy of the incoming projectiles. These kinetic energies are large enough that the binary collision approximation can be used, which treats the process as a sequence of isolated elastic collisions between the projectile and individual target atoms [50]. The most prominent features in the spectra are the single scattering peaks (SSPs), which ride on a background due to multiple scattering. A SSP corresponds to a projectile that experiences a single hard collision with a surface atom and backscatters directly into the detector, losing energy primarily through a classical elastic collision. Thus, the larger the mass of the target atom, the higher the scattered energy of the singly scattered projectile. Note that the time scale in the figure is reversed so that increasing energy of the scattered projectiles is towards the right. In Fig. 1, the Te SSP is at 4.7 μs and the Bi SSP is at 4.1 μs. The intensity of a given total yield SSP is basically proportional to the number of atoms that are directly visible to both the incoming ion beam and the detector multiplied by the differential scattering cross section at that scattering angle. In this case, the cross section for scattering from Te is about 1.5 times less than for scattering from Bi [51].

The relative orientations of the ion beam, sample and detector can be used to make the LEIS spectra reflect only particular atoms in a single crystal sample by aligning the incident beam and/or the detector along low index crystalline directions [26,52]. For example, a normally incident beam, i.e., along the [0001] direction of $Bi_2(Se,Te)_3$, will only directly impact the three outermost atomic layers, as the atoms beneath are completely shadowed by the upper layers when using ions in the low energy regime. Such an arrangement in which the beam is incident along a low index direction, but the detector is not, is called single alignment. In addition, the detector can also be positioned along a low index crystalline direction so that projectiles scattered from deeper lying atoms are blocked from reaching the detector by the outermost atoms, which is called double alignment. For $Bi_2Te_3$ surface, as indicated by the schematic diagram in the upper left panel of Fig.1, when the detector is along the ($1\bar{2}10$) azimuthal plane and positioned 50° from the surface normal, the projectiles scattered from 2nd layer Bi and 3rd layer Te atoms are completely blocked by the first layer Te atoms and single scattering is only possible from the first atomic layer [26,34]. The diagram includes a projection onto the surface plane of a representation of the blocking cone that prevents projectiles scattered from the 2nd and 3rd layers from reaching the detector. The same protocol is used for the $Bi_2Se_3$ surface except that the detector is positioned at 55º from the surface normal. The upper panels in Fig. 1 show spectra collected in both single and double alignment from clean $Bi_2Te_3$. In single alignment, which probes the composition of the outermost three atomic layers, both Te and Bi SSPs are visible. The Bi SSP is very small in double alignment but it is not absent, however, which is possibly due to surface defects and/or edge atoms at surface terraces. It can thus be concluded that the clean IBA-prepared $Bi_2Te_3$ surface is primarily Te-terminated, as is expected from the bulk crystal structure of this 2D layered material.

**B. Halogen exposure of Se- or Te-terminated bulk $Bi_2(Se,Te)_3$**

The middle and lower panels in Fig. 1 show TOF spectra collected from $Bi_2Te_3$ after exposure to $Br_2$ and then after post-annealing, respectively. The clean sample is exposed to 10 µA of $Br_2$ and TOF spectra are collected after each hour of exposure. The spectra do not change after 2 hrs of exposure, which means that the Br coverage has saturated. The middle panel of Fig. 1 shows spectra collected after 30 µA hrs of $Br_2$ exposure, i.e., the saturated surface. A peak at 5.5 µs that corresponds to the Br SSP appears after $Br_2$ exposure, and the intensities of the Te and Bi SSPs reduce in both alignments, which indicates that Br adsorbs on the surface and partially shadows or blocks the substrate atoms. Note that the cross section for scattering from Br is about 2.3 times less than for scattering from Bi [51], so that the small size of the Br SSP corresponds to more adsorbed Br than it may appear.

The coverages of all of the surface elements before and after halogen exposure to $Bi_2Te_3$ are summarized in Table 1. The SSP areas are calculated by integrating the peaks after subtracting the multiple scattering background, with the errors being determined by assuming that they are purely statistical, i.e., the square root of the total number of counts in each SSP. The coverages are calculated from the ratio of the Br SSP to Bi SSP after normalizing by the relative differential scattering cross-sections [51] and the MCP detection efficiency [48]. The coverages are calibrated using the assumption that the number of visible surface Te atoms in double alignment and the number of visible Bi atoms in single alignment from the as-prepared surfaces are each 1 monolayer (ML). Here, 1 ML is defined to be the number of atoms in the first layer of the defect-free TI surface. It is found that the coverage of Br on the saturated surface is about 0.57 ML, and the Te SSP attenuates by about 30% in both alignments, while the Bi SSP only decreases in single alignment. The clean surface has 0.1 ML of Bi visible in double alignment, which implies that the surface contains Te vacancy defects, which could be sites that Br adatoms attach to. The Bi SSP in double alignment has no decrease after adsorption, however, and the amount of adsorbed Br is larger than the number of defects that are present on the clean surface. Thus, it appears that Br does stick to the Te-terminated surfaces at sites other than defects. After Br adsorbs, the LEED pattern gets weaker and shows no higher order spots, indicating that the adsorbed Br atoms are not ordered and the surface itself may be somewhat disordered, possibly by the breaking of some Bi-Te bonds, which could be the reason why the Bi SSP in double alignment increases in intensity after Br exposure.

After reaching a saturation coverage of Br, annealing at 130°C for 2 hrs is performed to explore the surface reactions. The post annealing spectra in the bottom panels of Fig. 1 are very similar to those collected from clean bulk $Bi_2Te_3$, meaning that the Br SSP disappears and the Te and Bi SSPs recover their intensity. This indicates that the bonding between Br and the surface is weak and this low temperature annealing easily breaks the bonds. Br adatoms likely recombine into $Br_2$ as they desorb and any broken Bi-Te bonds reform, thus restoring the intact $Bi_2Te_3$ substrate.

As Cl has a much smaller atomic mass than Br, it cannot be detected directly by TOF due to the poor efficiency of the MCP in measuring particles with low kinetic energy. Table 1 shows that after $Cl_2$ exposure, the amounts of Bi and Te at the surface do not substantially change, as in the case of $Br_2$ exposure. This implies that many fewer Cl atoms stick to $Bi_2Te_3$ than do Br atoms. To confirm this, spectra were collected with the ESA for 3.0 keV $Na^+$ scattering at 40° along two different azimuths from $Bi_2Te_3$ and from the Ta sample holder after 40 μA hrs of exposure to $Cl_2$, as shown in Fig. 2. Note that neither of these orientations correspond to alignments along a low index crystalline direction, but the small scattering angle means that the beam and detector are close to the surface plane, so that most of the deeper lying atoms are shadowed by the surface atoms making this measurement primarily sensitive to the outermost couple of atomic layers. After $Cl_2$ exposure, a Cl SSP appears in the spectrum collected from the sample holder, which verifies that the sample has been exposed to $Cl_2$ and that this orientation is sensitive to the presence of Cl on the surface. The amount of Cl adsorbed on the sample holder is calculated to be roughly 0.5 ML. A Cl SSP is not, however, observed in spectra collected from the samples under either azimuthal alignment, which indicates that a significant amount of $Cl_2$ does not stick to the $Bi_2Te_3$ surface. In the same way, there is no Cl SSP visible in ESA spectra collected following $Cl_2$ reaction with Se-terminated $Bi_2Se_3$ surfaces (not shown). Note that the possibility that a small number of Cl atoms adsorb at vacancy defects cannot absolutely be excluded from these data, however, because under the grazing incident angles used in collecting the ESA spectra, Cl atoms adsorbed at a vacancy defect could be shadowed by surface Se or Te atoms. In addition, the differential cross section for 3.0 keV Na scattering at 40° from Cl is 1.65 times smaller than for Br [51], so that the measurement is less sensitive to small amounts of Cl. Finally, note that the TOF spectra collected after $Cl_2$ exposure and annealing are identical to those following $Br_2$ reaction, suggesting that if any chlorine were adsorbed that it recombines and desorbs under annealing thus restoring the clean surface in a similar manner as annealing a $Br_2$-reacted surface.

Additionally, for the ion scattering methods employed here, neither the TOF apparatus nor the ESA have sufficient energy resolution to distinguish the Br from the Se SSP due to their close atomic masses ($M_{Se}$ = 79 u, $M_{Br}$ = 80 u). Thus, the information that can be gleaned from the LEIS spectra is that Br does not adsorb at a site that shadows the Bi SSP in single alignment, implying that it does not adsorb to any great degree. Another measurement method is needed, however, to absolutely confirm that $Br_2$ does not stick to Se-terminated $Bi_2Se_3$.

### C. Halogen exposure of Bi/Bi$_2$(Se,Te)$_3$

Halogen exposures and annealing are also performed on Bi-capped $Bi_2(Se,Te)_3$ samples. Deposited Bi grows in a quasi bilayer-by-bilayer mode on $Bi_2(Se,Te)_3$ surfaces as Bi bilayers oriented along the (0001) direction, as they have a similar hexagonal lattice structure [53-55]. The first Bi BL has a relatively strong bonding to the $Bi_2Se_3$ surface and sits flat on both surfaces. Additional deposited Bi grows epitaxially on top of previous bilayers, but multilayers start to form before any particular bilayer fully covers the surface, which leads to films that are not uniformly thick. To ensure that the samples are covered with at least one Bi BL while not yet being bulk-like Bi, a slight excess of 1.5 Bi BLs is grown on the samples prior to halogen exposure [55].

It is found that $Br_2$ and $Cl_2$ always adsorb on the Bi BL-terminated $Bi_2(Se,Te)_3$ surfaces. Figure 3 shows TOF spectra collected from $Bi/Bi_2Te_3$ before and after $Br_2$ adsorption. All of the spectra are collected in single alignment and the upper curves are the total scattered yields. For the surface covered with a 1.5 Bi BL film, the spectra in the left panel show a prominent Bi SSP, as expected. There is still some Te SSP remaining, but it is much less than that observed from clean $Bi_2Te_3$ and is due to the fact that the ion beam and detector can still see $3^{rd}$ layer Te atoms in single alignment. It is also possible there are a small number of pinholes in the Bi film that expose some of the underlying Te-terminated surface. After the sample is reacted with 10 µA hrs of $Br_2$, the Br SSP appears and the Bi SSP intensity is reduced while the Te SSP intensity is unchanged, as seen in the right panel. This suggests that the Br attaches atop to the surface Bi atoms and not to any of the exposed Te atoms.

The number of visible Bi, Te and Br surface atoms before and after halogen adsorption is summarized in the lower three rows of Table 1. After a 10 µA hrs exposure of $Br_2$, more than 0.7 ML of Br atoms adsorb on surface, which is more than the amount that attaches to clean $Bi_2Te_3$. The large decrease of the Bi surface coverage shows that Br favors atop adsorption to Bi sites over Te. The same decrease of the Bi SSP intensity occurs for the $Cl_2$ exposed $Bi/Bi_2Te_3$, as seen in Table 1, and for halogen exposed $Bi/Bi_2Se_3$ surfaces (data not shown), which further illustrates how halogen atoms attach directly to Bi much more readily than to Te or Se.

Another useful feature of TOF spectra is that the site-specific neutral fraction (NF) can be obtained for each SSP, which provides information on the local electrostatic potential (LEP) directly above the scattering site [47,56,57]. According to resonant charge transfer (RCT) model typically used to model neutralization in low energy alkali ion scattering, when the projectile is close to a surface, the ionization level shifts and broadens while electrons tunnel between states in the surface and the ionization level. The measured neutralization probability thus depends on the relative positions of the projectile's ionization level and the surface Fermi energy at a "freezing distance", which is typically a few Å's above the scattering site [58], as the process is non-adiabatic due to the high velocity of the projectile. The NF for each SSP is calculated by dividing the integrated area of the neutral SSP by the area of total SSP after subtracting the multiple scattering background. The RCT model predicts that the NF in scattering from conventional solids decreases when the Fermi level moves down as the LEP (sometimes referred to as the local work function) increases, and vice versa, which has been confirmed by multiple experiments and calculations [56,57,59].

For halogens adsorbed on bulk $Bi_2(Se,Te)_3$ and $Bi/Bi_2(Se,Te)_3$ hetero-structures, the electronegative halogen adatoms are expected to behave as hole donors, i.e., as p-dopants. P-doping of an initially intrinsic semiconductor would normally increase the work function and consequently decrease the NF [60], but the neutralization in alkali LEIS depends on the surface potential and not necessarily on the bulk doping since the surface Fermi level could be pinned by surface states. In Fig. 3, the shaded areas show the TOF spectra of the scattered neutrals, and that ratio of the SSP areas in the neutral to the total yield spectra indicate the NFs. Table 2 summarizes the Bi SSP NFs in single alignment before and after $Br_2$ and $Cl_2$ exposure of clean $Bi_2(Se,Te)_3$, $Bi/Bi_2Se_3$ and $Bi/Bi_2Te_3$.

When clean $Bi_2Se_3$ and $Bi_2Te_3$ surfaces are exposed to halogens, the first three rows of Table 2 show very subtle decreases in the NF, which indicate a minor p-doping effect. The difference is too small compared to the relative large error limits, however, to be conclusive. One reason for such a small change is that halogens only loosely, or do not at all, adsorb on the Te- and Se-terminated surfaces, so that the LEP and thus the NF, do not change significantly. Additionally, clean $Bi_2Se_3$ and $Bi_2Te_3$ have high work functions that result in very low NFs, so that changes in the NF are difficult to discern.

When $Bi_2(Se,Te)_3$ is covered with a Bi BL, the Bi SSP NF increases from that of the clean surface because Bi has a lower work function than $Bi_2(Se,Te)_3$ and serves as an electron donor. Thus, the NF becomes more sensitive to changes in the LEP that occur after halogen exposure. In comparing the left and right panels of Fig. 3 and the data in the last three rows of Table 2, it is seen that after $Br_2$ exposure of $Bi_2Te_3$, both the Te and Bi NFs decline. When $Cl_2$ is reacted with a $Bi/Bi_2(Se,Te)_3$ hetero-structure, TOF spectra show no Cl peak, but there is a reduction of the intensity of the Bi SSP and the overall NF decreases in a similar manner as for $Br_2$ (data not shown). All of the Bi SSP NFs drop to the 6%-8% level following halogen adsorption. Note that because the Se and Te SSPs are small, it is difficult to get accurate values for their NFs, but it is clear from the spectra that they also decrease. Since the halogen atoms adsorb to Bi sites, it can be assumed that they form local downward pointing dipoles by attracting electrons from the surface Bi atoms, thus increasing the work function. The NF data thus confirm that these dipoles increase the LEP at the Bi sites, which then decreases the measured NFs, as is expected for p-doping [61].

In addition, a special LEED pattern is found when a surface prepared by deposition of 1.5 Bi BLs on $Bi_2Se_3$ is exposed to 6.7 μA hrs of $Br_2$, as shown in Fig. 4. Note that the background is brighter on the left side of the image due to inhomogeneities in the phosphor screen. The pattern consists of higher order six-fold spots in the center, which means that the Br on the surface forms a $\sqrt{3} \times \sqrt{3}$ $R30°$ superstructure. The LEED pattern stays the same when more $Br_2$ is reacted, indicating that the surface Br coverage is saturated by this point. The higher order LEED pattern is only found on the $Br_2/Bi/Bi_2Se_3$ system, as the other Bi BL-covered surfaces all maintain a well ordered 1x1 LEED pattern with the Bi BL crystal periodicity following halogen adsorption.

### D. Annealing of halogen-exposed surfaces

A 130°C low temperature anneal is performed on halogen-exposed $Bi_2(Se,Te)_3$ samples to study any surface chemical reactions. It is found that the TOF spectra have no change in either peak intensity nor NF after annealing. This is either because the halogens do not stick to the surfaces, or that they are so weakly bonded to the Se or Te surface atoms that a low temperature heating causes the halogen atoms to desorb leaving the Se or Te-terminated sample surfaces intact.

For Bi-covered surfaces, however, etching of the Bi atoms occurs after annealing. Figure 5 shows double alignment TOF spectra collected from $Bi_2Te_3$ after Bi is deposited, following exposure to $Br_2$, and then after it is post-annealed. The upper spectrum shows mostly the Te SSP, confirming the Te-termination. The next spectrum was collected after deposition of 1.5 Bi BLs, which causes a Bi SSP to appear. There's still some Te SSP visible, which is likely due to an incomplete coverage of Bi. A control experiment is then performed by annealing the Bi-capped sample at 130°C for 3 hrs in the absence of Br, as indicated by the dashed curve. In this case, the TOF spectra do not show any decrease of the Bi SSP intensity nor an increase of the Te SSP, which excludes the possibility of surface Bi being removed by annealing alone.

The $Bi/Bi_2Te_3$ surface is then exposed to 10 μA hrs of $Br_2$, which leads to the appearance of a Br SSP and the attenuation of the Te and Bi SSPs, as seen in the 4$^{th}$ spectrum of Fig. 5. The attenuation of the Bi is greater than in Fig. 3 as only the top layer atoms are detected in double alignment, which implies that the Br attaches mainly to the top layer Bi atoms. Then, after annealing $Br_2$-reacted $Bi/Bi_2Te_3$ at 130°C for 2 hrs, the Br SSP disappears and the Bi SSP is smaller than in the TOF spectra collected from the original Bi-capped sample. After an additional cycle of $Br_2$ exposure and annealing, as shown in the bottom spectrum of Fig. 5, the Bi SSP is very small and the Te SSP gets even larger, becoming close in area to that of clean $Bi_2Te_3$. This indicates that Bi atoms are removed during the annealing, presumably through an etching process in which Br reacts to form the volatile $BiBr_3$ compound, or $BiCl_3$ in the case of chlorine adsorption. This etching reaction removes the surface Bi and reveals the Te-terminated surface of the bulk sample.

Figure 6 is a summary of the proportion of surface Bi remaining following halogen adsorption onto and annealing of samples covered with Bi films, shown as a function of halogen exposure. Different annealing times are used in each cycle for the different reactants and materials, as these are determined by the time that it takes for the TOF spectra to saturate, i.e., display no further changes after additional annealing. The proportion of Bi is calculated as the ratio of the Bi SSP area to the total Bi and Se SSP areas. Note that the difference in differential cross section for Na scattering from Bi and Se (or Te) and MCP efficiency for scattered projectiles with different kinetic energies is included in calculating the total areas. The amount of surface Bi remaining declines with the amount of halogen exposure, and eventually reaches the value of pristine $Bi_2(Se,Te)_3$ bulk materials. It is thus found that the Bi films on $Bi_2(Se,Te)_3$ are etched away by a thermally driven reaction with adsorbed Br or Cl and that the removal of Bi is nearly complete with a sufficient halogen exposure.

**IV. Discussion**

According to the LEIS spectra, a limited number of Br atoms stick to clean Te-terminated $Bi_2Te_3$, but no evidence of Cl adsorption on either QL-terminated $Bi_2Se_3$ or $Bi_2Te_3$ is found. This is unexpected as $Cl_2$ is normally more reactive than $Br_2$, but it is partially because Cl atoms have a smaller LEIS cross-section than Br due to its smaller mass making Cl more difficult to discern in the spectra. In addition, it is possible that some Cl is located beneath the top layer, such as at a defect site so that it is directly attached to a Bi atom, which would cause the Cl to be shadowed by surface Te or Se atoms.

The bulk-terminated TI surfaces remain intact after annealing, either because halogens do not adsorb or they are only weakly bonded. The limited bonding of molecular halogens to clean TIs may be partially due to these surfaces being relatively inert two-dimensional van der Waals materials, or because the terminating Se or Te atoms are unreactive because they have filled valence electron orbitals and are thus unable to break a halogen-halogen bond. The fact that Se- and Te-terminated surfaces are basically inert also means that halogens do not replace the surface atoms and bond to second layer Bi atoms. Thus, for clean and defect-free $Bi_2(Se,Te)_3$, it is difficult to modify the surface structure with halogen molecules. This is unlike the behavior of group V-terminated III-V semiconductor surfaces, for which halogens are able to dissociate and break the bonds between atoms in the first two layers and attach to the reactive group III atoms in the second layer [22,23].

The notion that Br can occupy Se or Te surface vacancy defect sites is worth exploring, however, as the aging effect is believed to result in n-doping from such vacancies, so that filling the vacancies with an electronegative halogen atom may help to compensate for the missing surface atoms and restore the Fermi level position. For the samples used in the present work, however, the number of surface vacancy defects is small and cannot account even for the small amount of Br that adsorbs on Te-terminated $Bi_2Te_3$ surfaces. Thus, this is still an open question and a future experiment is planned in which vacancy defects are intentionally produced to determine if halogen do occupy such sites, and whether they act as p-dopants when they do.

For Bi-covered surfaces, however, halogen adsorption readily occurs. There are a couple of reasons why this is likely. First, although Bi is also a two-dimensional van der Waals material, the manner in which it grows on $Bi_2(Se,Te)_3$ is not through a pure layer-by-layer process [53-55] so that there are many defects, step edges, etc., at which the $Br_2$ or $Cl_2$ can readily react and dissociatively adsorb. Second, due to Bi being more electropositive than Se or Te, it can more easily give up electrons to induce the dissociative adsorption of halogen molecules. This is parallel to the behavior of group III-terminated III-V semiconductors in which the surface atoms have empty surface states that readily react with halogens [22,24]. There is a large decrease of the NF of the Bi SSP with halogen adsorption implying that the bonding Bi atoms donate charge to Cl or Br, which is expected as Bi is more metallic and halogens are more electronegative. This implies a decrease of the Fermi level position and that halogens are a good candidate for p-doping of Bi-covered heterostructures. Further confirmation of this hypothesis can be obtained from angle-resolved photoelectron spectroscopy (ARPES), transport or other measurements.

In addition, etching of the halogen-covered Bi films occurs when the samples are heated. This is similar to the processes that occur with various other semiconductor surfaces that are etched by adsorbed chlorine and bromine [2,3]. For silicon reacted with $Cl_2$, for example, SiCl and $SiCl_2$ form on the surface and annealing a Cl-saturated surface to about 700 K removes the surface Si atoms in the form of gaseous silicon chlorides. For III-V semiconductors, such as GaAs reacted with $Cl_2$, volatile chlorides of both Ga and As form and the material is etched when the sample is heated at ~650 K [3]. A similar reaction can be inferred to occur between surface Bi and halogens. Halogens initially adsorb dissociatively to the surface Bi atoms. When heated, Bi then reacts with the halogens to form volatile molecules, which are most likely the stable Bi trihalides. The melting point of $BiCl_3$ is 227°C and 219°C for $BiBr_3$, which are relatively low, and the vapor pressures of $BiCl_3$ and $BiBr_3$ at 130°C are about $3\times10^{-3}$ Torr [62]. Thus, when the halogen-covered samples are heated to only 130°C, the halides evaporate leaving bare Se- or Te-terminated surfaces.

## V. Conclusions

Halogen adsorption and reaction with $Bi_2Se_3$, $Bi_2Te_3$ and $Bi/Bi_2(Se,Te)_3$ surfaces is systematically studied by LEIS and LEED. $Br_2$ can weakly bond to Te-terminated $Bi_2Te_3$ surfaces, while $Cl_2$ shows no evidence of adsorption to either $Bi_2Se_3$ or $Bi_2Te_3$. The halogens that do stick desorb after a light annealing, leaving the surface intact and with the same structure as that of the initially prepared clean surface, which further demonstrates the inertness of QL-terminated TIs. Halogens do readily adsorb on Bi BL films on TI substrates, however, and they have a p-doping effect. The Bi bilayers are etched by adsorbed halogens when heated at a relatively low temperature, which provides the possibility of fabricating nanoscale Bi structures through a controlled etching process, in addition to other methods such as direct bottom-up growth [40] and focused ion beam nanofabrication [63].

Heterostructures consisting of a small number of Bi layers deposited onto TI or other materials are attracting more interest and such a controlled halogen chemical reaction suggests a way to manipulate the amount of Bi on the surfaces. The atomic level etching process requires further study, however, such as determining if the etching occurs in a layer-by-layer mode and whether the etching starts from the edge or if it is isotropic. Future studies from our group will focus on the effects of halogen adsorption on vacancy defect sites, especially monitoring any electronic structure changes such as Fermi level shifting.

## VI. Acknowledgements

This material is based on work supported by the U.S. Army Research Laboratory and the U.S. Army Research Office under Grant No. 63852-PH-H.

Bi(111) Bilayer on Sb Nanofilms by Quantum Confinement Effect", ACS Nano **10**, 3859-64 (2016).

[40] Z. F. Wang, M.-Y. Yao, W. Ming, L. Miao, F. Zhu, C. Liu, C. L. Gao, D. Qian, J.-F. Jia, and F. Liu, "Creation of helical Dirac fermions by interfacing two gapped systems of ordinary fermions", Nat. Comm. **4**, 1384 (2013).

[41] F. Yang, L. Miao, Z. F. Wang, M.-Y. Yao, F. Zhu, Y. R. Song, M.-X. Wang, J.-P. Xu, A. V. Fedorov, Z. Sun, G. B. Zhang, C. Liu, F. Liu, D. Qian, C. L. Gao, and J.-F. Jia, "Spatial and energy distribution of topological edge states in single Bi(111) bilayer", Phys. Rev. Lett. **109**, 016801 (2012).

[42] L. Miao, Z. F. Wang, W. Ming, M.-Y. Yao, M. Wang, F. Yang, Y. R. Song, F. Zhu, A. V. Fedorev, Z. Sun, C. L. Gao, C. Liu, Q.-K. Xue, C.-X. Liu, F. Liu, D. Qian, and J.-F. Jia, "Quasiparticle dynamics in reshaped helical Dirac cone of topological insulators", PNAS **110**, 2758–62 (2013).

[43] T. Hirahara, G. Bihlmayer, Y. Sakamoto, M. Yamada, H. Miyazaki, S. Kimura, S. Blugel, and S. Hasegawa, "Interfacing 2D and 3D topological insulators: Bi(111) bilayer on $Bi_2Te_3$", Phys. Rev. Lett. **107**, 166801 (2011).

[44] I. R. Fisher, M. C. Shapiro, and J. G. Analytis, "Principles of crystal growth of intermetallic and oxide compounds from molten solutions", Philos. Mag. **92**, 2401-35 (2012).

[45] W. Zhou, H. Zhu, and J. A. Yarmoff, "Spatial distribution of topological surface state electrons in $Bi_2Te_3$ probed by low-energy $Na^+$ ion scattering", Phys. Rev. B **97**, 035413 (2018).
26

**Table 1.** Coverage of surface atoms calculated from ratios of the single scattering peak areas in LEIS TOF spectra collected in both double and single alignment orientations from Bi$_2$Te$_3$ and 1.5 Bi BL covered Bi$_2$Te$_3$ surfaces before and after exposure to Br$_2$ and Cl$_2$. No data is presented for chlorine coverages as the Cl SSP is not visible in the TOF spectra.

|  | Double Alignment | | | Single Alignment | | |
| --- | --- | --- | --- | --- | --- | --- |
|  | Br | Te | Bi | Br | Te | Bi |
| **Bi$_2$Te$_3$ as prepared** |  | 1±0.06 | 0.10±0.01 |  | 1.6±0.1 | 1±0.04 |
| **30 μA hrs Br$_2$ exposure** | 0.56±0.07 | 0.74±0.06 | 0.16±0.02 | 0.58±0.06 | 1.05±0.05 | 0.66±0.03 |
| **30 μA hrs Cl$_2$ exposure** |  | 0.98±0.09 | 0.07±0.01 |  | 1.47±0.15 | 0.91±0.09 |
| **Bi covered Bi$_2$Te$_3$** |  | 0.34±0.04 | 0.89±0.04 |  | 0.43±0.04 | 1.52±0.05 |
| **10 μA hrs Br$_2$ exposure** | 0.8±0.1 | 0.22±0.04 | 0.26±0.03 | 0.68±0.06 | 0.47±0.04 | 1.03±0.04 |
| **10 μA hrs Cl$_2$ exposure** |  | 0.44±0.07 | 0.31±0.04 |  | 0.40±0.06 | 0.93±0.07 |

**Table 2.** The neutralization probability for 3.0 keV Na$^+$ singly scattered from Bi in a single alignment orientation from clean Bi$_2$Se$_3$, Bi$_2$Te$_3$ and these samples covered with 1.5 BL of Bi before and after 10 µA hrs exposures to Br$_2$ and Cl$_2$.

|  | Bi$_2$Se$_3$ | Bi$_2$Te$_3$ |
|---|---|---|
| **Clean IBA prepared surface** | 4.9% ± 0.4% | 6.3% ± 0.7% |
| **After Br$_2$ exposure** | 4.7% ± 0.5% | 6.0% ± 0.5% |
| **After Cl$_2$ exposure** | 3.1% ± 0.3% | 5.3% ± 0.9% |
| **1.5 Bi BL covered surface** | 43% ± 1% | 34% ± 1.1% |
| **After Br$_2$ exposure** | 7.1% ± 0.7% | 6.8% ± 0.5% |
| **After Cl$_2$ exposure** | 6.6% ± 0.5% | 8.3% ± 1.1% |

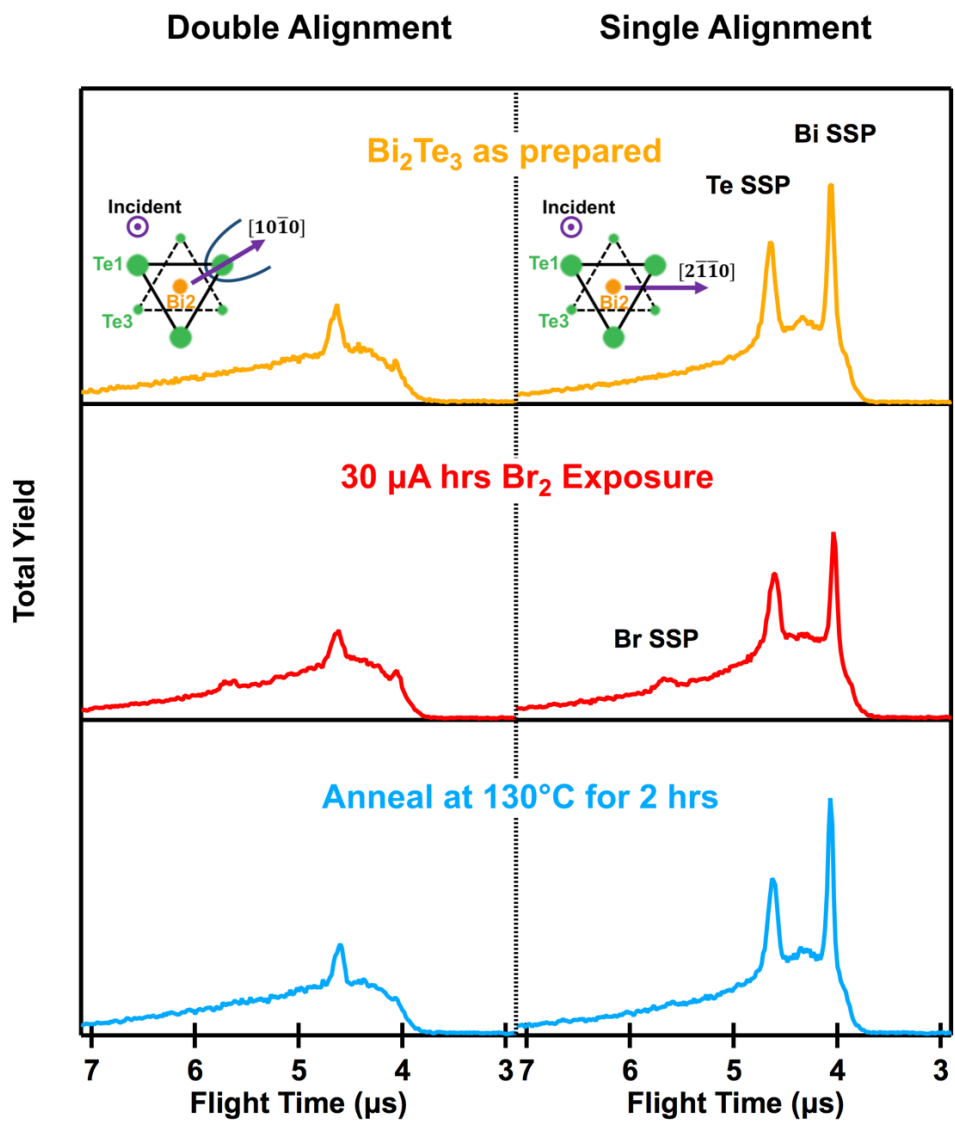

**Figure 1.** LEIS total yield TOF spectra collected from (top) clean $Bi_2Te_3$, (middle) after 30 μA hrs of $Br_2$ exposure, and (bottom) post-annealing at 130 °C for 2 hrs. The 3.0 keV $Na^+$ ion beam is incident normal to sample surface and the detector is positioned at a scattering angle of 130°. The upper panels include top view schematics with projections of the azimuthal directions of the scattered beams, the atoms and their layer numbers indicated. The spectra in the left panel are collected along the $(1\bar{2}10)$ azimuthal plane, which is a double alignment configuration. The spectra in the right panel are collected in single alignment by rotating the azimuthal angle with respect to the detector by 30°. The intensities of the spectra are normalized to each other using the ion beam current measured on the sample and the data collection time.

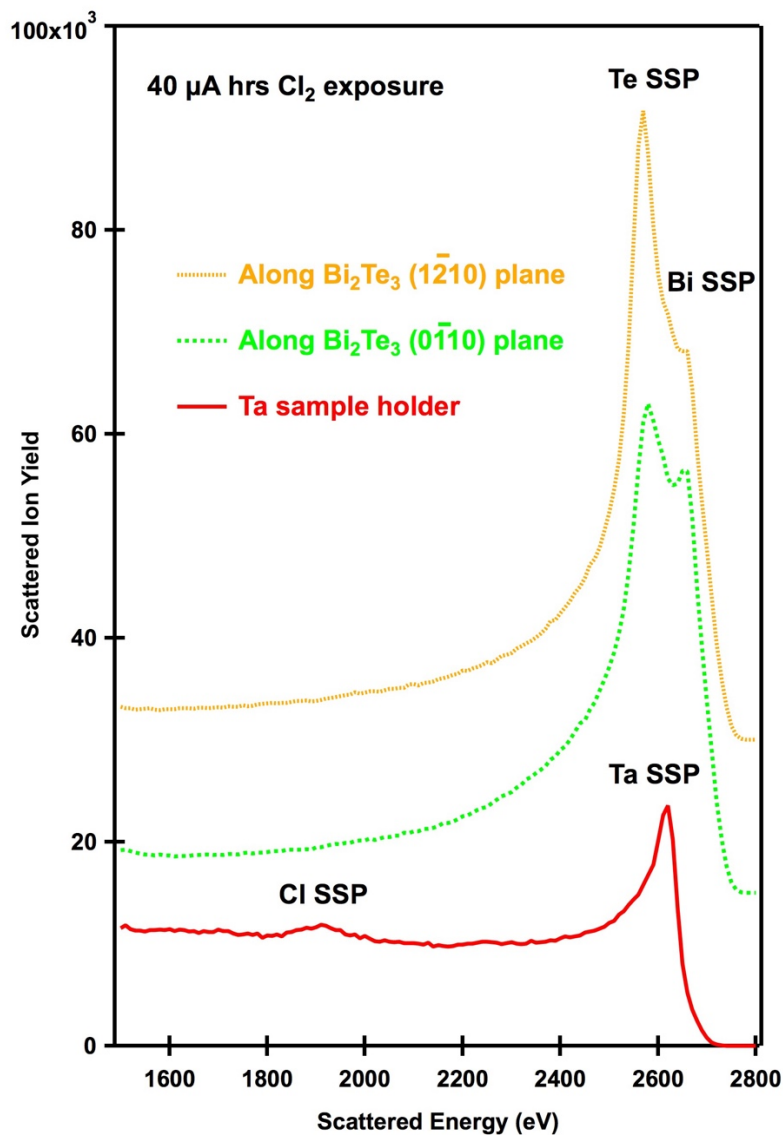

**Figure 2.** 3.0 keV Na⁺ LEIS spectra collected with the ESA from $Bi_2Te_3$ and the Ta sample holder following 40 µA hrs of $Cl_2$ exposure. The beam is incident at a 70° angle from the surface normal and the detector is positioned at a scattering angle of 40°. The spectra are offset vertically for clarity.

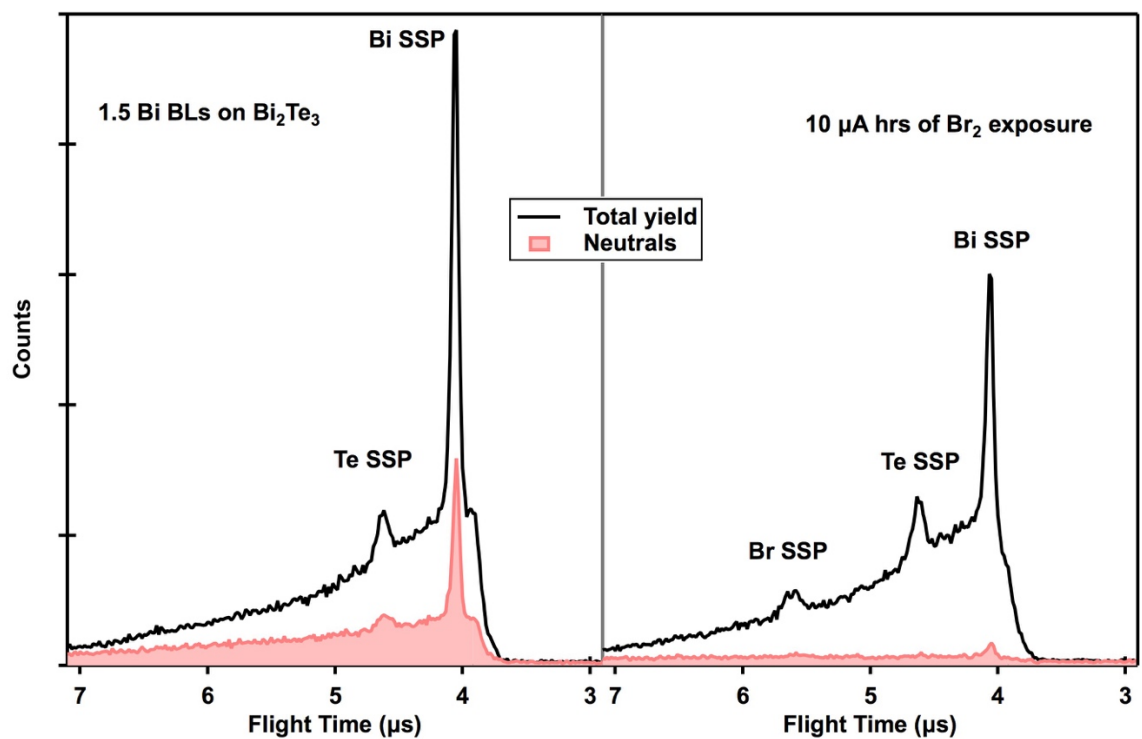

**Figure 3.** 3.0 keV Na$^+$ LEIS TOF spectra collected from surfaces of (left) 1.5 Bi BL covered Bi$_2$Te$_3$, and (right) after exposure to 10 μA hrs of Br$_2$. All of the spectra are collected in single alignment. The solid black lines show the total yield spectra of scattered projectiles, while shaded areas show spectra of the scattered neutral particles.

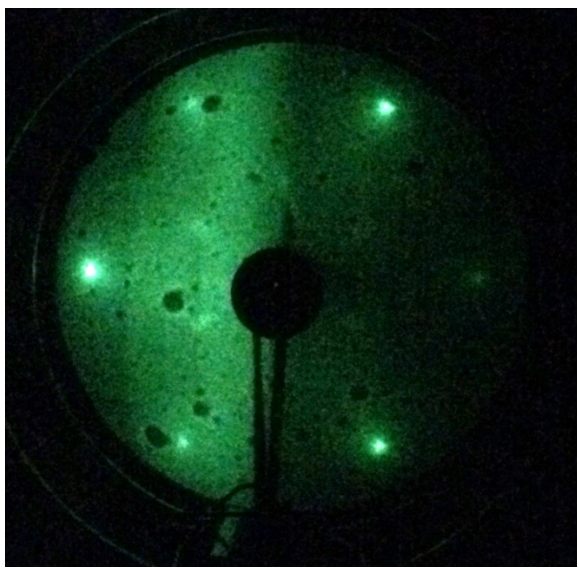

**Figure 4.** LEED pattern collected from 1.5 Bi BL covered Bi$_2$Se$_3$ exposed to 6.7 µA hrs of Br$_2$. The image is collected using a beam energy of 21.8 eV, and shows a $\sqrt{3} \times \sqrt{3}\ R30°$ LEED pattern.

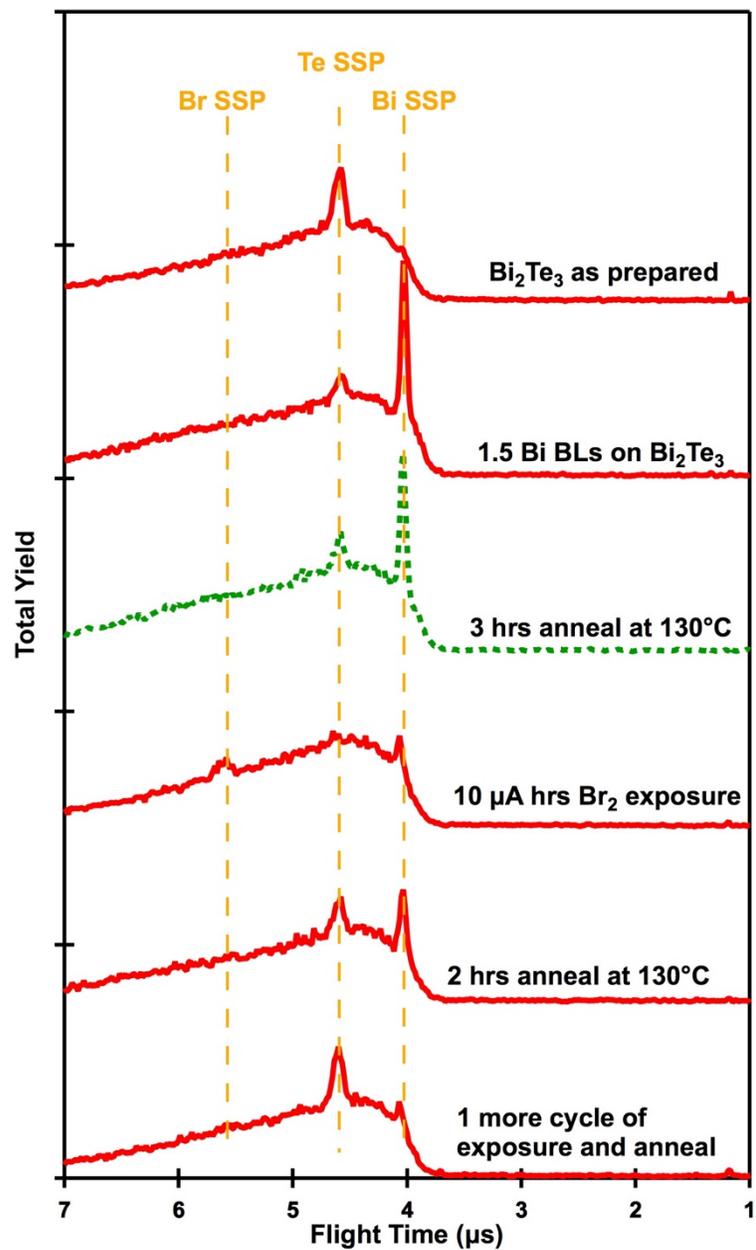

**Figure 5.** 3.0 keV Na$^+$ LEIS TOF spectra collected in double alignment from the surfaces of clean Bi$_2$Te$_3$ and after subsequent treatments as marked in the graph and described in the text.

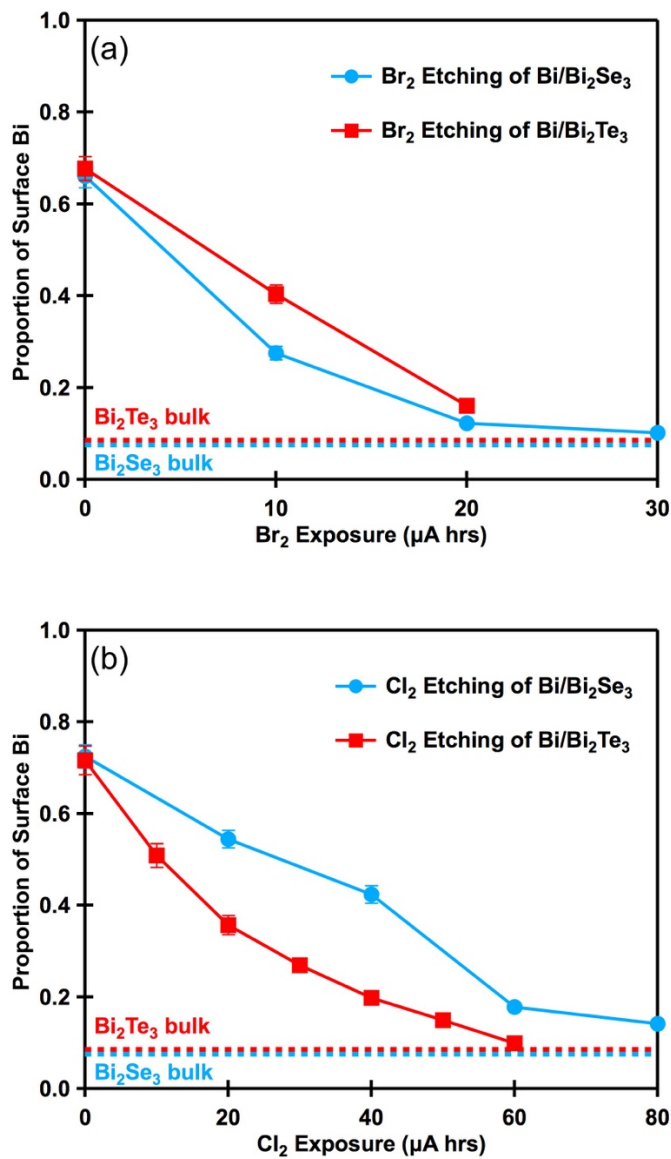

**Figure 6.** The proportion of Bi on the surface following cycles of halogen adsorption and annealing shown as a function of (a) $Br_2$ exposure and (b) $Cl_2$ exposure. The annealing is performed at a temperature of 130°C for 4 hrs with $Br_2$ etching of Bi/$Bi_2Se_3$, 2 hrs for $Br_2$ etching of Bi/$Bi_2Te_3$, 0.5 hrs for $Cl_2$ etching of Bi/$Bi_2Se_3$ and 2 hrs for $Cl_2$ etching of Bi/$Bi_2Te_3$. The horizontal dashed lines indicate the proportions measured from the clean bulk-terminated surfaces. Note that the error bars associated with many of points are smaller than the symbols.